# THE DAEδALUS PROJECT: RATIONALE AND BEAM REQUIREMENTS

Jose R. Alonso, for the DAEδALUS Collaboration
MIT, Cambridge, MA 02138

*Abstract*

Neutrino physics focuses on huge detectors deep underground. The Sanford Lab in South Dakota will build a 300 kiloton water-Cherenkov detector 1500 meters deep for muon neutrino oscillation studies of the mass hierarchy and CP violation. This will be used by the Long Baseline experiment (LBNE) detecting few GeV neutrinos from Fermilab, 1300 km away. The DAEδALUS Collaboration also plans several neutrino-production sites at closer distances up to 20 km from the 300 kT detector, producing muon antineutrinos from stopped pions. The complementarity with LBNE greatly enhances results, and enthusiasm is mounting to do both experiments. DAEδALUS needs 0.8-1 GeV accelerators with mA proton beams. Three sites at 1.5, 8 and 20 km from the 300 kT detector require several accelerators. The cost per machine must be below 1/10 of existing megawatt-class proton machines. Beyond high power and energy, beam parameters are modest. Challenges are reliability, control of beam loss and minimizing activation. Options being studied are: a compact superconducting cyclotron; a ring cyclotron accelerating $H_2^+$ (with stripping extraction); and a stacked cyclotron with up to 9 planes sharing the same magnet yoke and rf systems.

## INTRODUCTION

Neutrinos are very much at the forefront of fundamental physics today. The discovery of oscillation between neutrino flavors ($\nu_e$, $\nu_\mu$, $\nu_\tau$) – implying mass for the neutrino, and allowing the potential for CP-violation studies – has vaulted the neutrino into a position of prominence as a tool for exploration of and possibly for extension beyond the Standard Model. Many experiments are being mounted, based on very large, highly-sensitive detectors, located mainly in deep underground areas where rock overburden provides shielding from cosmic muons.

A new detector in this category is being planned for the DUSEL project sited at the 4850 Level of the Sanford Lab at the Homestake mine in Lead, South Dakota. This detector is planned to consist of 300 kilotons of Gd-doped water as a Cherenkov counter, and will serve as the Far detector in the Long Baseline Neutrino Experiment (LBNE) for a neutrino beam produced 1300 km away at Fermilab.

The DAEδALUS Collaboration is proposing to use this same detector, and to mount several high-power accelerators, as sources of antineutrinos from stopped π+, at varying distances from the target [1]. This configuration provides sensitivity to a CP-violating term in the coupling matrix that is quite measurable within the DAEδALUS configuration alone, but the combination of DAEδALUS and LBNE data can be used to perform a measurement of $\delta_{CP}$ substantially better than the measurement from either LBNE or DAEδALUS alone.

Key to the DAEδALUS experiment is a suitable flux of antineutrinos, which translates into megawatt-class sources of protons at energies of the order of 800 MeV. The planned configuration for the experiment calls for three sites, at 1.5 km, 8 km and 20 km from the detector. In addition, as multiple accelerators will be needed (possibly more than one at the farther sites), cost considerations are extremely important.

High-power beams are of extreme interest in numerous fields, from high-level waste transmutation to driving subcritical reactors, for spallation neutron sources to isotope production and materials interrogation schemes for national security applications. Now, neutrino sources can be added to the list.

At this conference, several technology alternatives will be presented, indicating that the requirements for the DAEδALUS experiment are not outside the realm of technical feasibility. This is particularly true as the timetable for the experiment is tied to the availability of the DUSEL large detector, probably not on line before 2021.

Nonetheless, the technological and funding challenges should not be underestimated, so concentrated efforts addressing both challenges are needed to ensure that a suitable set of accelerators is available when neutrinos are needed for the measurements.

## NEUTRINO OSCILLATIONS AND THE STANDARD MODEL

In the simplest version of the theory of particle physics, positing three quark and three lepton families transitions, should involve only conversions between members of the same family. Weak decays of hadrons, however, are more complicated, showing transitions forbidden by this rule, implying mixing in the final states between members of different families (Figure 1). For a long time it was thought that lepton families would obey the rule, however the discovery of oscillation between neutrino flavors clearly implied mixing in lepton families as well.

Mixing can be expressed as a rotation in coordinate systems, with the degree of mixing given as the angle of rotation (Figure 2). Figure 3 summarizes the oscillation process, where the critical parameters are $\sin^2 2\theta$, expressing the amplitude of oscillation between neutrino types, and $\lambda$, the oscillation wavelength, which is proportional to $E/\Delta m^2$, the energy of the neutrino divided by the mass difference squared between the two neutrino mass eigenstates. Note, the presence of oscillations clearly indicates that at least two of the neutrino mass eigenstates have nonzero mass. While measuring oscillation parameters provides a mass difference between neutrino types, there is no straightforward way of determining the absolute neutrino masses from oscillation measurements alone.

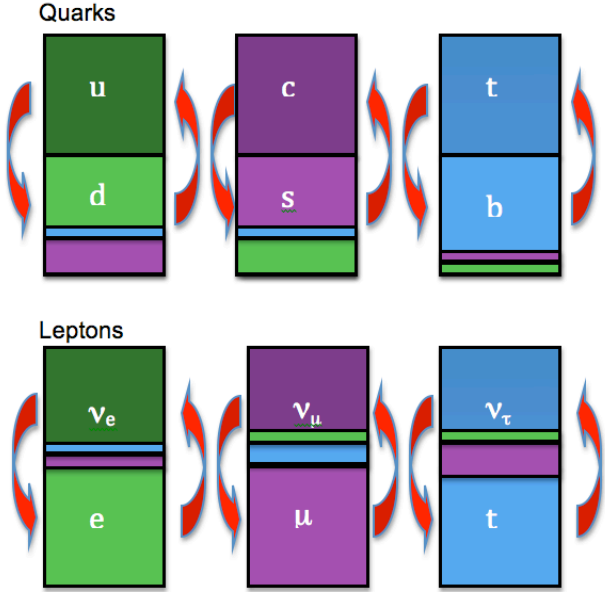

Fig. 1: Standard Model, indicating mixing in both quark and lepton families (from Ref. [2])

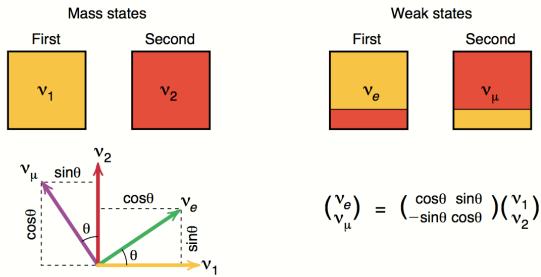

Fig. 2: Mixing in 2-state system, expressed as a rotation through an angle θ.

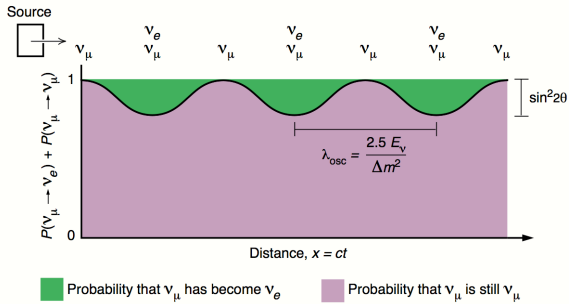

Fig. 3: Oscillation of ν (moving to right) between $\nu_\mu$ and $\nu_e$ states. Amplitude of oscillation is related to $\sin^2 2\theta$, wavelength λ related to the energy of the neutrino $E_\nu$ divided by the (mass difference)$^2$ between the two states.

Mixing is a complex process. The three neutrino flavors, $\nu_e$, $\nu_\mu$, $\nu_\tau$ couple into the three neutrino "mass eigenstates" $\nu_1$, $\nu_2$ and $\nu_3$ by a matrix U in the form of Eq. (1).

$$\begin{pmatrix} \nu_e \\ \nu_\mu \\ \nu_\tau \end{pmatrix} = \begin{pmatrix} U_{e1} & U_{e2} & U_{e3} \\ U_{\mu 1} & U_{\mu 2} & U_{\mu 3} \\ U_{\tau 1} & U_{\tau 2} & U_{\tau 3} \end{pmatrix} \begin{pmatrix} \nu_1 \\ \nu_2 \\ \nu_3 \end{pmatrix} \quad (1)$$

The matrix elements are combinations of the mixing angles $\sin\theta_{ij}$ $\cos\theta_{ij}$ ($s_{ij}$, $c_{ij}$) and the CP-violating term $\exp(-i\delta)$ parameters, expanded in Eq. (2). Appropriate manipulations result in Eq (3) expressing the probability of a muon neutrino oscillating into an electron neutrino.

$$U = \begin{bmatrix} c_{12}c_{13} & s_{12}c_{13} & s_{13}e^{-i\delta} \\ -s_{12}c_{23}-c_{12}s_{23}s_{13}e^{i\delta} & c_{12}c_{23}-s_{12}s_{23}s_{13}e^{i\delta} & s_{23}c_{13} \\ s_{12}s_{23}-c_{12}c_{23}s_{13}e^{i\delta} & -c_{12}s_{23}-s_{12}c_{23}s_{13}e^{i\delta} & c_{23}c_{13} \end{bmatrix} \quad (2)$$

$$\begin{aligned} P = \quad & (\sin^2\theta_{23}\sin^2 2\theta_{13})(\sin^2\Delta_{31}) \\ \mp & \sin\delta\,(\sin 2\theta_{13}\sin 2\theta_{23}\sin 2\theta_{12})(\sin^2\Delta_{31}\sin\Delta_{21}) \\ + & \cos\delta\,(\sin 2\theta_{13}\sin 2\theta_{23}\sin 2\theta_{12})(\sin\Delta_{31}\cos\Delta_{31}\sin\Delta_{21}) \\ + & (\cos^2\theta_{23}\sin^2 2\theta_{12})(\sin^2\Delta_{21}). \end{aligned} \quad (3)$$

In this equation the terms $\Delta_{ij} = \Delta m_{ij}^2\, L/4E\nu$, where L is the baseline distance between source and detector. The most sensitive portion of the CP-violating term ($\sin\delta$) has a sign difference for neutrino and antineutrino oscillations.

The Standard Model provides a framework for physical processes, for instance allowing for mixing between families, however it does not make predictions for the important parameters involved in these processes. These are determined by experiment. Many of the parameters governing behavior in this sector of particle physics have been measured experimentally: $\theta_{12}$ determined from solar neutrino measurements (disappearance of $\nu_e$); $\theta_{23}$ from atmospheric neutrino measurements (disappearance of $\nu_\mu$), and the respective mass differences. Table 1 summarizes these experimentally measured numbers.

Table 1 (from Ref. [1])

| Parameter | Present: Value | Uncert. (±) |
|---|---|---|
| $\Delta m_{21}^2 \times 10^{-5}$ eV$^2$ | 7.65 | 0.23 |
| $\Delta m_{31}^2 \times 10^{-3}$ eV$^2$ | 2.40 | 0.12 |
| $\sin^2(2\theta_{12})$ | 0.846 | 0.033 |
| $\sin^2(2\theta_{23})$ | 1.00 | 0.02 |
| $\sin^2(2\theta_{13})$ | 0.06 | 0.04 |

Remaining to be determined is an accurate measurement for $\theta_{13}$. (Current 1-sigma error bar is ~100% of value.) Experiments are underway with precision measurements of reactor-produced antineutrinos which are expected to provide a value for θ₁₃ within about 5 years (Daya Bay and Double Chooz).

## CP VIOLATION MEASUREMENTS

The traditional method for measuring the CP-violating term, $\delta_{CP}$, is to look at the difference between oscillation probabilities for muon neutrinos and antineutrinos, using the + and – signs of the second term in Eq 3. This is the basis for the CP-violation measurements derived from the long-baseline experiments being mounted or in operation around the world: T2K; Minos and NOνA; CERN-to-Gran Sasso (ICARUS, OPERA); and the LBNE at the Sanford Laboratory (Homestake-DUSEL). Interactions of high intensity proton beams produce pions, which are focused by a strong solenoidal horn that, by polarity reversal, alternatively selects $\pi^+$ and $\pi^-$. The pions are allowed to decay in a long channel, the resulting $\nu_\mu$ or $\overline{\nu_\mu}$ being directed to the near and far detectors of the experiment. Measuring the appearance of $\nu_e$ and $\overline{\nu_e}$ in the far detector provides the oscillation rate. The unknown parameters are $\delta_{CP}$ and $\theta_{13}$ in the relevant equation.

There is another way of evaluating the CP-violating term, which is to use the L dependence of the interference ($\Delta_{31}$ and $\Delta_{21}$) terms in Eq. 3, and to perform measurements at different distances. The DAEδALUS experiment plans to exploit the second method of measuring $\delta_{CP}$. By producing the same neutrino spectra at three carefully-chosen distances from the detector, a sensitive determination of $\delta_{CP}$ can be made.

Note that although there are many terms in the expression of Eq. 3, Table 1 shows good numerical values for all but the $\theta_{13}$ mixing angle, and the CP-violating term $\delta_{CP}$. Sensitivity of experiments is usually expressed on the so-called "jelly-bean plot", Fig. 4, which shows a contour of uncertainty around a potential location within the $\sin^2\theta_{13}$, $\delta_{CP}$ space. The collection of "jelly beans" in Figures 7-9 traces the size and shape of these uncertainties for a particular experiment in different regions of this $\sin^2\theta_{13}$, $\delta_{CP}$ space. The smaller the "jelly bean" the smaller the uncertainty in the measured point.

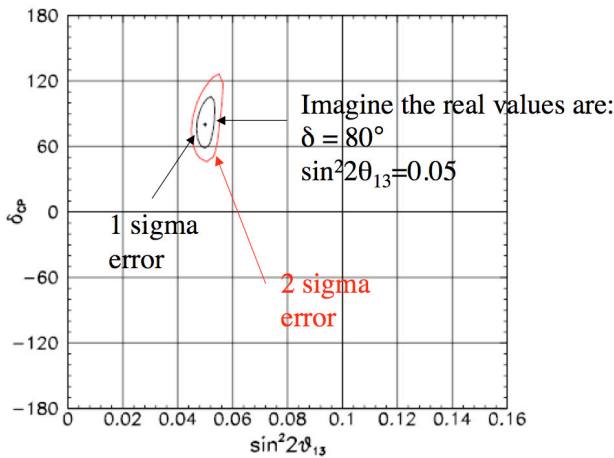

Fig. 4: Sample of "Jelly-Bean" plot

## THE DAEδALUS EXPERIMENT

To be sited around the Sanford Lab, the DAEδALUS experiment uses the same 300 kton water Cherenkov detector planned for LBNE. DAEδALUS will place neutrino sources at three locations, 1.5 km (on the surface directly above the detector), and 8 km and 20 km from the detector (Figure 5).

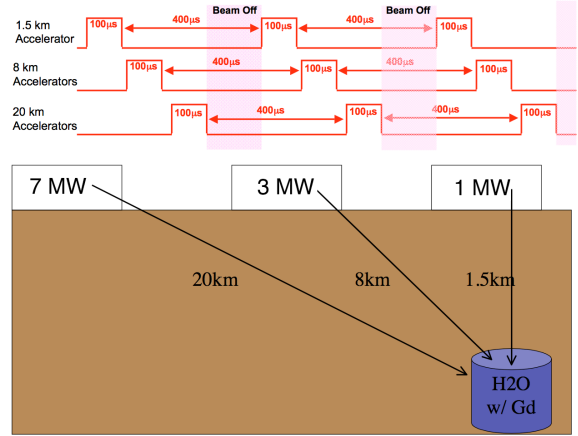

Fig. 5: Schematic of DAEδALUS experiment. Three accelerator complexes at 1.5, 8 and 20 km distance from the detector, with approximate beam powers of 1 MW, 3 MW and 7 MW respectively, produce adequate neutrino fluxes for the $\delta_{CP}$ measurements. Each accelerator runs at a staggered times, each with 20% duty factor, enabling tagging of neutrino events in the detector by time of arrival. The beam-on time is arbitrary, but greater than 100 μs.

Each of the neutrino sources will require sufficient power in its proton beam to generate a suitable flux of neutrinos at the detector site. This translates into beam energies around 1 GeV and power levels in the megawatt range. Such protons striking a target produce a large number of relatively low-energy pions. If the target is large enough these pions will be slowed down and stopped. In the process almost all of the $\pi^-$ will be captured by target nuclei, while each $\pi^+$ will be stopped resulting in its subsequent decay to a $\mu^+$ and $\nu_\mu$, and a few microseconds later to a positron accompanied by a $\overline{\nu_\mu}$ and $\nu_e$ pair. The energy spectra of the various components are shown in Figure 6.

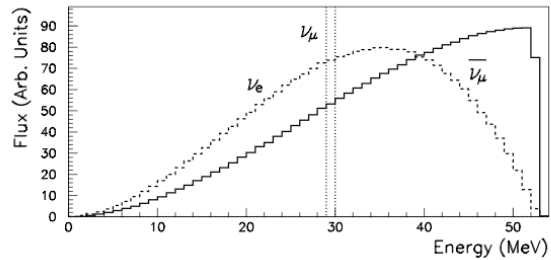

Fig. 6: Energy spectrum of neutrinos from a "decay-at-rest" beam.

Of note is that there are no ($\overline{\nu_e}$)s produced at a significant level. (These would result from decay of $\pi^-$, but as these are all absorbed before they can decay, the resulting neutrino spectra from these proton beams are devoid of $\overline{\nu_e}$.)

Thus, if in a detector one observes ($\overline{\nu_e}$)s, their appearance must come from oscillations of ($\overline{\nu_\mu}$)s.

A detector with a lot of protons (such as the planned water-Cherenkov LBNE detector) is sensitive to $\overline{\nu_e}$, and through an Inverse Beta Decay (IBD) interaction the antineutrino is captured by a proton, releasing a positron and a free neutron. This event produces a unique signature in the detector: the prompt slowing and annihilation of the positron, and a delayed capture of the neutron. This delayed coincidence enables discrimination against many of the background processes occurring in the detector.

As described above, carefully selecting the distance L of the neutrino sources can maximize difference between oscillation probabilities, and hence give a good measure of the CP-violating term in Eq. 3. Timing shown in Fig 5 enables tagging of events seen in the detector with the production site. Each accelerator is switched on for 20% of the time, with a 40% time when all beams are off to get a measure of background events. The beam-on time is arbitrary, with a minimum of about 100 μs to be substantially longer than the muon lifetime (2.2 μs), but other considerations such as thermal stability or beam-loading could play in the optimization of the cycle times. Only hard requirement is that each proton source be on for 20% of the time. This has the consequence that peak power must be 5 times higher than average power.

Modeling of the DAEδALUS experiment after running for a 10 year period yields the Jelly-Bean plot shown in Fig 7.

## COMPARISON OF DAEδALUS AND THE FERMILAB LONG BASELINE NEUTRINO EXPERIMENT (LBNE)

The LBNE experiment can run simultaneously with DAEδALUS. There is a short, sub-microsecond burst from Fermilab that can be gated out of the DAEδALUS beam structure, so it is not even necessary to synchronize timing between the two experiments. By having data-taking occurring simultaneously and in the same detector, systematic errors should be substantially if not completely reduced from the data-comparison work.

One can ask whether the long baseline LBNE and short baseline DAEδALUS can really be compared. Recall, though, that the critical parameter in the oscillation wavelength is the ratio of $L/E\nu$. For LBNE, L is 1300 km, $E\nu$ in the 10's of GeV range, while for DAEδALUS, L is around 10 km while $E\nu$ is in the 10's of MeV range. While both are a factor of 1000 different, their ratios are comparable. So, the answer to the feasibility of comparison is a resounding "yes."

Figure 8 shows the Jelly-Bean plot for LBNE, which is quite comparable to that of DAEδALUS. However, the most noteworthy analysis is shown in Fig. 9 where the data sets of LBNE and DAEδALUS are combined. Joining these data sets significantly improves the evaluation of $\delta_{CP}$, as shown by the significantly smaller jelly beans. This serendipitous result arises from the strengths of each technique playing to compensate for the weaknesses in the other.

As a result of this analysis, there is a growing excitement in the neutrino community for mounting both long- and short-baseline experiments.

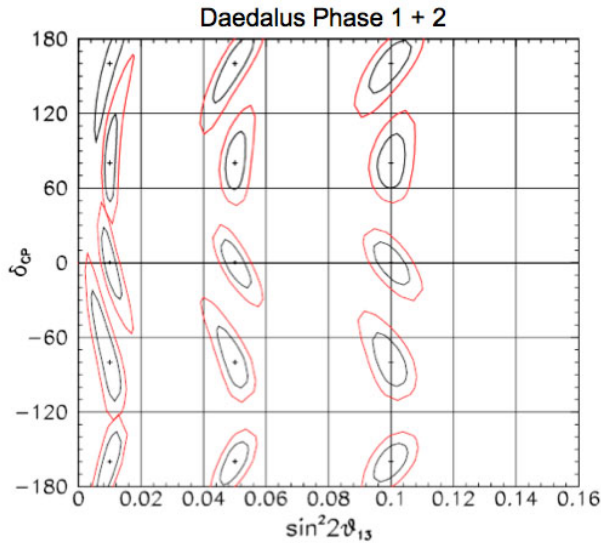

Fig. 7: DAEδALUS Jelly-Bean plot

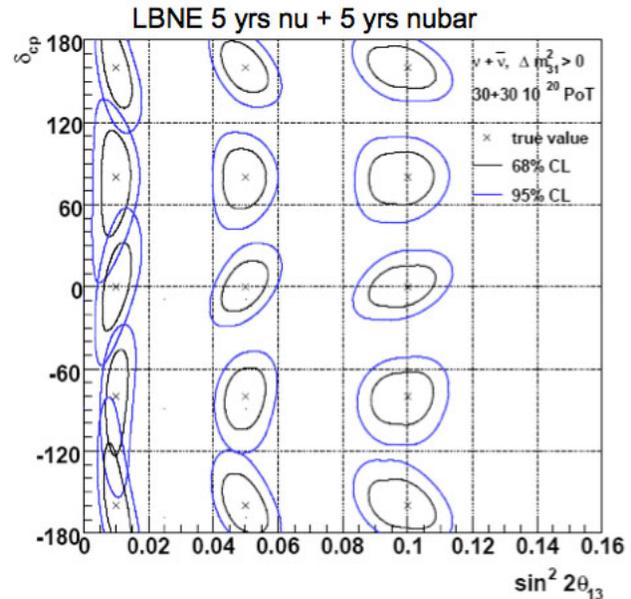

Fig. 8: LBNE Jelly-Bean plot

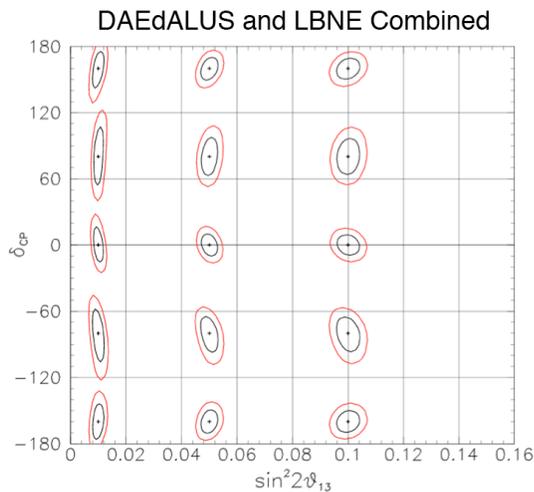

Fig. 9: Jelly-Bean plot for combined LBNE and DAEδALUS data sets. Quality of $\delta_{CP}$ measurement is substantially improved!

## ACCELERATOR REQUIREMENTS

Table 2 summarizes the rather general accelerator requirements. No special beam qualities are needed, just raw power at the appropriate energy within a broad time window. The other important considerations are reliability: the machines must run with adequate up-time for a 10 year period, and moderate to low acquisition and operations cost.

Table 2: Basic accelerator parameters

| Parameter | Requirement |
| --- | --- |
| Energy | 1 GeV ± 400 MeV |
| Power-Average | 1-7 MW |
| Power-Peak | 5-35 MW |
| | may need > 1 machine at each site |
| Duty factor | 20% each site |
| Beam-on time | >100 μs |
| Beam loss | <0.1% for acceptable activation and maintainability |
| Cost | Low <$100M/machine |

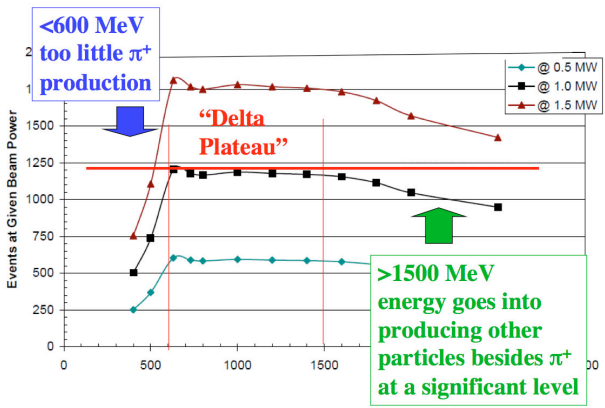

Fig. 10: π production vs proton energy showing flat rate between 600 MeV and 1.5 GeV.

Figure 10 indicates that pion production over the energy range between 600 MeV and 1.5 GeV is pretty flat, coming from the Delta resonance. However below 600 MeV proton energy, i.e. below Delta resonance energies, pion-production efficiency drops precipitously. An appropriate minimum beam energy would be around 800 MeV to optimize production in a thick target.

Figure 5 shows the timing requirements, again beam-on times for the experiment can be fairly arbitrary (as long as greater than 100 μs), so can be driven by accelerator or target engineering considerations. The powers shown for each station in Figure 5 are average powers, so beam currents during beam-on time must reflect instantaneous powers 5 times higher. As neutrinos are emitted from pions and muons at rest, they are isotropic in space. Hence the flux at the detector will vary as $1/r^2$. However, the sensitivity of the oscillation increases at the longer distance, so less statistics are needed to obtain a meaningful measurement of the oscillation probability. This very favorable condition substantially reduces the power required at the far station. So, the 1/3/7 MW (average) power values adequately express the requirements of the experiment.

The most challenging accelerator requirement is minimization of beam loss. At these energies and powers, beam loss leads to component activation and in many cases to outright destruction of parts. The rule of thumb for linacs is: allowed beam loss of 1 watt/meter will still enable hands-on maintenance. For compact machines, with component sizes of the order of a few meters, this would indicate beam losses of $10^{-6}$. While this is impractical, and hence hands-on maintenance will be impossible, critical components will need to be designed to handle the highest possible heat load. In any event, for a megawatt beam, overall efficiency must be of the order of 99.9% (total beam loss less than $10^{-3}$).

## TECHNOLOGY OPTIONS

The list of applications of GeV/megawatt-class beams is quite extensive, and now should include neutrino sources. Currently the world's highest power facility is PSI, with 2.2 mA of 590 MeV protons producing 1.3 MW of beam power. Superconducting linacs, such as the SNS at Oak Ridge, are close competitors. Both, however, represent very high capital investments. The challenge for the DAEδALUS project is to obtain the energy/power beams at a fraction of the cost of the above facilities. The task is not hopeless: substantial innovations have brought new technologies into or close to readiness for application.

Noteworthy is the $H_2^+$ design study being conducted at INFN Catania for a superconducting ring cyclotron with stripping extraction. Reported by Luciano Calabretta at this conference [3], the concept shows great promise towards meeting the project requirements.

Other options are: a compact high-field concept proposed by Timothy Antaya [4], and a stacked cyclotron concept being developed by Peter McIntyre from Texas A&M [5].

Superconducting linac designs are also on the table; and would probably present less technical risk, however the high cost and substantially larger footprint make this option less attractive at this stage.

## PROSPECTS AND SUMMARY

The timetable for the DAEδALUS experiment allows a brief cushion for development and selection of the best accelerator technology, as it requires completion of the LBNE detector at Homestake. Current estimates are that this detector will not be available before about 2021. However, ten years will pass very quickly, and it will take a concerted effort to meet the technical challenges presented in the accelerator design. The technology field is wide open at the present time, and we invite, nay urge active participation by the broad accelerator community in addressing these challenges.

## ACKNOWLEDGMENTS

The author wishes to express gratitude to Janet Conrad (MIT) for support, both financial and technical.

## REFERENCES

[1] "Expression of Interest for A Novel Search for CP Violation in the Neutrino Sector: DAEdALUS": June, 2010: arXiv.org > physics > arXiv:1006.0260
[2] Los Alamos Science #25 (1997): http://la-science.lanl.gov/lascience25.shtml
[3] L. Calabretta, "A Multi Megawatt cyclotron complex to search for CP violation in the neutrino sector," Cyclotrons-10, Lanzhou, Sept. 2010, TUA1CIO01, http://www.JACoW.org.
[4] T. Antaya, "Energetic Cyclotrons as Devices – Pushing beyond 6 Tesla," Cyclotrons-10, Lanzhou, Sept. 2010, TUA1CIO02, http://www.JACoW.org.
[5] G. Kim, D. May, P. McIntyre, A. Sattarov: "A Superconduction Isochronous Cyclotron Stack as a Driver for a Thorium-Cycle Power Reactor," Cyclotrons 01, East Lansing, May 2001, http://www.JACoW.org